\documentclass[floatfix,twocolumn,aps,prd,showpacs,amsmath,amssymb,draft]{revtex4}

\usepackage{epsfig}
\usepackage{color}

\begin{document}

\title{
Quantization of the Proca field in the Rindler wedge and 
the interaction of uniformly accelerated currents with
massive vector bosons from the Unruh thermal bath. 
}

\author{Jorge Casti\~neiras}
\email{jcastin@ufpa.br}
\affiliation{Faculdade de F\'\i sica, Universidade Federal do
Par\'a, 66075-110, Bel\'em, Par\'a, Brazil}

\author{Emerson B. S. Corr\^ea}
\email{emerson@ufpa.br}
\affiliation{Faculdade de F\'\i sica, Universidade Federal do
Par\'a, 66075-110, Bel\'em, Par\'a, Brazil}

\author{Lu\'\i s C. B. Crispino}
\email{crispino@ufpa.br}
\affiliation{Faculdade de F\'\i sica, Universidade Federal do
Par\'a, 66075-110, Bel\'em, Par\'a, Brazil}

\author{George E. A. Matsas}
\email{matsas@ift.unesp.br}
\affiliation{Instituto de F\'\i sica Te\'orica, Universidade 
Estadual Paulista, Rua Dr. Bento Teobaldo Ferraz, 271-Bl. II, 
01140-070 S\~ao Paulo, S\~ao Paulo, Brazil}

\date{\today}

%%%%%%%%%%%%%%%%%%%%%%%%%%%%%%%%%%%%%%%%%%%%%%%%%%%%%%%%%%%%%%%%%%%%%%%%%%%%
\begin{abstract}
We canonically quantize the Proca field in the Rindler wedge and compute the 
total response rate of a uniformly accelerated current interacting with massive 
vector Rindler particles from the Unruh thermal bath. We explicitly 
verify that the result obtained is exactly the same as the emission rate 
of massive vector particles in the Minkowski vacuum as analyzed by inertial 
observers. Eventually our results are interpreted in terms of the interaction 
of static electrons coupled to $Z^0$ bosons present in Hawking radiation close 
to the event horizon of a black hole.  
\end{abstract}
\pacs{03.70.+k, 04.62.+v}
%%%%%%%%%%%%%%%%%%%%%%%%%%%%%%%%%%%%%%%%%%%%%%%%%%%%%%%%%%%%%%%%%%%%%%%%%%%%

\maketitle

%%%%%%%%%%%%%%%%%%%%%%%%%%%%%%%%%%%%%%%%%%%%%%%%%%%%%%%%%%%%%%%%%%%%%%%%%%%%
\section{Introduction}
\label{Introduction}
%%%%%%%%%%%%%%%%%%%%%%%%%%%%%%%%%%%%%%%%%%%%%%%%%%%%%%%%%%%%%%%%%%%%%%%%%%%%

According to the Unruh effect~\cite{U} (see also Refs.~\cite{F,BD}), 
the vacuum state of a quantum field theory as described by inertial 
observers in Minkowski spacetime corresponds to a thermal state as 
seen by uniformly accelerated ones confined to the Rindler wedge
(Rindler observers). By now, it is thoroughly accepted the essential 
role played by the Unruh effect in the description of various physical 
phenomena~\cite{CHM-RMP}. The analysis 
of the total response rate for classical and semiclassical sources 
with constant proper acceleration from the point of view of Rindler and
inertial observers has been already performed for Klein-Gordon~\cite{eric,CSM}, 
Maxwell~\cite{HMS-campo de Maxwell} and Dirac fields~\cite{MV}. 

Here we canonically quantize the Proca field (i.e., a massive 
spin-1 vector field) in the Rindler wedge and compute the response 
rate for a source with constant proper acceleration coupled to it.  
We explicitly verify that the total response rate as computed by 
Rindler observers immersed in the Unruh thermal bath is precisely 
the same as the one calculated by inertial observers in the Minkowski 
vacuum. Because  uniformly accelerated sources are static in Rindler 
coordinates, they can only interact with zero-energy Rindler particles 
of the Unruh thermal bath. (We recall that this is possible because 
massive Rindler particles may have arbitrarily small frequencies as
defined by uniformly accelerated observers~\cite{CCMV}.)
Under proper conditions, our results can be interpreted
in terms of noninertial electrons by saying that each 
$Z^0$ emitted from a uniformly accelerated $e^-$ in the Minkowski 
vacuum as described by inertial observers corresponds to either 
the emission to or absorption from the Unruh thermal bath of a 
zero-energy $Z^0$ Rindler boson as described by coaccelerated 
observers. 

The paper is organized as follows. In Sec.~\ref{CQ}, the Proca field is 
quantized in the Rindler wedge. In Sec.~\ref{source}, we briefly revisit 
the regularization procedure applied to static sources in the Rindler 
wedge as originally derived in Ref.~\cite{HMS-campo de Maxwell}. In 
Sec.~\ref{RF} the corresponding emission and absorption rates of 
zero-energy Rindler particles are computed. In Sec.~\ref{IF} it is 
explicitly verified that by combining the previously computed emission 
and absorption rates as calculated by Rindler observers, we exactly 
obtain the emission rate of (nonzero-energy) Minkowski particles as 
computed by inertial observers. In Sec.~\ref{Conclusions}, we make our 
final remarks and comment on the response of a static charge interacting 
with $Z^0$ bosons of Hawking radiation in the vicinity of a black hole. 
We adopt natural units $\hbar = c = G = 1$, unless stated otherwise.

%%%%%%%%%%%%%%%%%%%%%%%%%%%%%%%%%%%%%%%%%%%%%%%%%%%%%%%%%%%%%%%%%%%%%%%%%%%
\section{Canonical quantization of the Proca field in the Rindler wedge}
\label{CQ}
%%%%%%%%%%%%%%%%%%%%%%%%%%%%%%%%%%%%%%%%%%%%%%%%%%%%%%%%%%%%%%%%%%%%%%%%%%%
In order to quantize the Proca field we start with the
standard Lagrangian density
\begin{eqnarray}
{\cal L} = \sqrt{-g}\left(-\frac{1}{4}F_{\mu\nu}F^{\mu\nu} + 
\frac{1}{2}m^2A_{\mu}A^{\mu} 
\right),
\label{Lagrangiana de Proca}
\end{eqnarray} 
where $g $ is the determinant of the metric $g_{\mu \nu}$
and 
$F^{\mu \nu}\equiv \nabla^{\mu}A^{\nu}-\nabla^{\nu}A^{\mu}$.
The field equations are
\begin{eqnarray}
\nabla_{\mu}F^{\mu \nu} + m^2A^{\nu} = 0. 
\label{Proca equation}
\end{eqnarray}
By applying $\nabla_\nu$ in Eq.~(\ref{Proca equation}), we obtain 
the Lorenz constraint
\begin{equation}
\nabla_{\mu}A^{\mu} =0.
	\label{Lorenz constraint}
\end{equation}
Equation~(\ref{Lorenz constraint}) can be used to cast the field 
equation~(\ref{Proca equation}) in the form
\begin{equation}
(\nabla^{\mu}\nabla_{\mu} + m^2)A^{\nu} = 0,
 \label{Lagrange equations}
\end{equation}
provided that the spacetime is a vacuum solution of 
Einstein equations: $R_{\mu \nu}=0$. 

Now, let us proceed to quantize the Proca field in the (right) 
Rindler wedge, i.e., the portion of the Minkowski spacetime 
defined as $z > |t|$, which is a globally hyperbolic spacetime by 
its own right. Here $(t,{\bf x})$ with ${\bf x} \equiv (x,y,z)$ are the 
usual Cartesian coordinates. We cover the Rindler wedge using coordinates 
$(\tau,x,y,\xi)$, where the Rindler coordinates $\tau$ and $\xi$ are 
implicitly defined through
\begin{equation}
t = \frac{e^{a\xi}}{a}\sinh{a\tau},\; 
z = \frac{e^{a\xi}}{a}\cosh{a\tau}
\end{equation}
and $a = {\rm const} \in {\bf R}_+$. Rindler observers will be 
uniformly accelerated ones with constant coordinates $x,y,\xi$, 
and corresponding proper acceleration $a e^{-a \xi}$.
By using Rindler coordinates, the metric components of the spacetime 
become
$$
g_{\mu\nu}= {\rm diag} \left( e^{2a\xi}, -1, -1, -e^{2a\xi} \right).
$$

In order to quantize the Proca field in the Rindler wedge,
it is convenient to find a complete set of orthonormal solutions
$A^{(\lambda,\varpi,{\bf k}_\bot)}$ for Eq.~(\ref{Lagrange equations}) 
satisfying the Lorenz constraint~(\ref{Lorenz constraint})
associated with the three possible physical polarizations 
$\lambda={\rm I},\, {\rm II},\, {\rm III}$:
\begin{equation}
(A^{(\lambda,\varpi,{\bf k}_\bot)},
A^{(\lambda^{\prime},\varpi^{\prime}, {\bf k}_\bot^{\prime})} ) 
=  
\delta_{\lambda \lambda'} 
\delta(\varpi-\varpi^{\prime})
\delta^2 ({\bf k}_\bot-{\bf k}_\bot^{\prime}),
\label{Rindler orthonormality}
\end{equation}
where  
${\bf k}_\bot \equiv (k_{x} , k_{y})$ represents the $z$-orthogonal momentum,
and $\varpi \in (0,+\infty)$ labels the Rindler frequency. (We recall that $\varpi$
is distinct from the frequency $\omega \in (m, +\infty)$ of Minkowski 
modes.) For this purpose, we consider the generalized Klein-Gordon inner product 
\begin{equation}
(A^{(i)},A^{(j)}) = \int_{\Sigma}d\Sigma \, n_{\mu}\,W^\mu (A^{(i)}, A^{(j)}),
\label{KG}
\end{equation}
where $(i),(j)$ stand for the set of quantum numbers $(\lambda, \varpi, {\bf k}_\bot)$, 
the Cauchy surface $\Sigma$ will be chosen to be any $\tau = {\rm const}$ hypersurface 
in the Rindler wedge, $n^\mu \propto (1,0,0,0)$ is the  future-pointing unit vector field 
orthogonal to  $\Sigma$, and 
\begin{equation}
W^\mu (A^{(i)}, A^{(j)}) \equiv
i ( \overline{A_{\nu}^{(i)}} \pi^{(j)\mu\nu} 
- {A^{(j)}}_{\nu} \overline{ \pi^{(i) \mu\nu}}),
\label{currentW}
\end{equation}
where the overline denotes complex conjugation, and
\begin{eqnarray}
\pi^{(i) \mu \nu}
& \equiv & 
\frac{1}{\sqrt{-g}} \frac{\partial \cal{L}}{\partial(\nabla_{\mu}A^{(i)}_{\nu})} 
\nonumber \\
& = &
\nabla^{\nu} A^{(i) \mu}-\nabla^{\mu} A^{(i) \nu}
\nonumber \\
& = &
- F^{(i) \mu \nu}.
\end{eqnarray}
By using Eq.~(\ref{Proca equation}) it is easy to show that 
$ W^\mu (A^{(i)}, A^{(j)}) $
is a conserved current: $\nabla_\mu W^\mu (A^{(i)}, A^{(j)})=0$.

A complete set of orthonormal solutions [i.e., complying with
Eq.~(\ref{Rindler orthonormality})] for Eq.~(\ref{Lagrange equations}) 
satisfying the constraint~(\ref{Lorenz constraint}) can be cast 
in the form
\begin{eqnarray}
&&
A_{\mu}^{({\rm I}, \varpi, {\bf k}_\bot)} = 
C^{({\rm I}, \varpi, {\bf k}_\bot)} ( 0, 0, k_y \phi , -k_x \phi ), 
\label{complete set of orthonormal solutions 1}
\\
&&
A_{\mu}^{({\rm II}, \varpi, {\bf k}_\bot)} = 
C^{({\rm II}, \varpi, {\bf k}_\bot)} \left( \partial_\xi \phi , 
-i \varpi \phi, 0 , 0 \right), 
\label{complete set of orthonormal solutions 2}
\\
&&
A_{\mu}^{({\rm III}, \varpi, {\bf k}_\bot)} = 
C^{({\rm III}, \varpi, {\bf k}_\bot)}  \nonumber \\
&& 
\times \left( - \frac{i \varpi k_\bot}{m} \phi, 
\frac{k_\bot}{m} \partial_\xi \phi, 
\frac{i k_x \rho^2}{m k_\bot} \phi, 
\frac{i k_y \rho^2}{m k_\bot} \phi \right),  
\label{complete set of orthonormal solutions 3}
\end{eqnarray}
where
\begin{eqnarray}
C^{({\rm I},\varpi,{\bf k}_\bot)} & = & 
\sqrt{\sinh(\pi\varpi/a)}/(2\pi^{2}k_\bot \sqrt{a}), \\
C^{({\rm II},\varpi,{\bf k}_\bot)} & = & 
\sqrt{\sinh(\pi\varpi/a)}/(2\pi^{2} \rho \sqrt{a}),  \\ 
C^{({\rm III},\varpi,{\bf k}_\bot)} & = & 
\sqrt{\sinh(\pi\varpi/a)}/(2\pi^{2} \rho \sqrt{a}).  
\end{eqnarray}
Here
\begin{equation}
	\phi (x^{\alpha})  \equiv 
	K_{i{\varpi}/{a}} ({\rho}e^{a\xi}/a )
	e^{i(k_{x}x+k_{y}y-\varpi\tau)}, \nonumber
\end{equation}
where
$\rho \equiv \sqrt{k_{\bot}^{2}+m^{2}}$,
$k_{\bot} \equiv |{\bf k}_\bot | $, 
and 
$K_{i {\varpi}/{a}}({\rho}e^{a\xi}/a)$ is the 
Bessel function of imaginary order~\cite{GR}.
It is interesting to note also that it does exist 
a forth mode: 
$$ 
A_{\mu}^{({\rm IV}, \varpi, {\bf k}_\bot)} = 
\frac{\sqrt{\sinh(\pi\varpi/a)}}{(2\pi^{2} m \sqrt{a})}
( -i \varpi \phi, \partial_\xi \phi, i k_x \phi , i k_y \phi )
$$
with negative norm,
$$
(A^{({\rm IV},\varpi,{\bf k}_\bot)},
 A^{({\rm IV},\varpi^{\prime}, {\bf k}_\bot^{\prime})} ) = 
 - 
 \delta(\varpi-\varpi^{\prime})
 \delta^2 ({\bf k}_\bot-{\bf k}_\bot^{\prime}),
$$ 
orthogonal to all $A_{\mu}^{(\lambda, \varpi, {\bf k}_\bot)}$
($\lambda = {\rm I, II, III}$). Although it satisfies 
Eq.~(\ref{Lagrange equations}), it does not comply with 
the constraint~(\ref{Lorenz constraint}) and thus must be 
considered nonphysical. This reflects the fact that 
the Proca field has only three independent physical 
polarizations.

Next, we expand the Proca field in terms of the physical modes
as 
\begin{eqnarray}
\hat{A}_{\mu}(x^{\nu}) &=& 
\int  d^2 {\bf k}_\bot 
\int_0^{\infty} d\varpi \sum_{\lambda={\rm I, II, III }} \nonumber \\
& & 
[\hat{a}_{(\lambda, \varpi, {\bf k}_\bot)}
A_{\mu}^{(\lambda, \varpi, {\bf k}_\bot)} + {\rm H.c.}],
\label{expansao do campo}
\end{eqnarray}
where 
$ 
\int d^2 {\bf k}_\bot \equiv 
\int_{-\infty}^{+\infty}dk_{x} \int_{-\infty}^{+\infty}dk_{y}
$.
In order to determine the commutation relations between the
creation 
$\hat{a}^{\dagger}_{(\lambda, \varpi, {\bf k}_\bot)}$ 
and annihilation 
$\hat{a}_{(\lambda, \varpi, {\bf k}_\bot)}$ 
operators, we introduce equal-time canonical commutation 
relations: 
\begin{equation}
 [ \hat{A}_{i}(x),\hat{A}_{j}(x^{\prime}) ]_\Sigma 
= 
[ \hat{\Pi}^{i}(x),\hat{\Pi}^{j}(x^{\prime} )]_\Sigma 
= 0,
\label{comutador}
\end{equation}
\begin{equation}
[ \hat{A}_{i} (x),\hat{\Pi}^{j}(x^{\prime} )]_\Sigma 
 = 
i\frac{\delta_i^j }{\sqrt{- g^{(3)}}}
\delta (\xi- \xi') \delta^2 ({\bf x}_\bot - {{\bf x}'}_\bot)
\label{comutador a com pi}
\end{equation}
on the Cauchy surface $\Sigma$ [covered with  coordinates $(\xi, {\bf x}_\bot)$ 
with ${\bf x}_\bot \equiv (x,y)$] among the field operators $\hat{A}_{i}$ and the 
canonically conjugate momenta $\hat{\Pi}^{j} \equiv n_{\mu} \hat{\pi}^{\mu j}$. 
Here $i,j $ label space coordinates and Eqs.~(\ref{comutador})-(\ref{comutador a com pi})
are cast in a noncovariant form because ${\hat \Pi}^{0} = n_\mu \hat \pi^{\mu 0}$ 
vanishes identically and $\hat A_0$ is not an independent variable 
(see, e.g., Ref.~\cite{GreinerR}).
Now, on the one hand by using Eq.~(\ref{expansao do campo}) we obtain 
\begin{eqnarray}
&& [(A^{(i)}, \hat A),(\hat A , A^{(j)} ))] 
\nonumber \\
&& =
\sum_{l p} 
(A^{(i)}, A^{(l)})
[\hat{a}_{{(l)}}, \hat{a}^{\dagger}_{(p)}] 
(A^{(p)}, A^{(j)})
\label{apoio1}
\end{eqnarray}
(where the sums above should be read as integral symbols 
for continuous quantum numbers).
On the other hand, by using Eqs.~(\ref{comutador})-~(\ref{comutador a com pi}), 
we have
\begin{equation}
[(A^{(i)}, \hat A),(\hat A , A^{(j)} ))]
=
(A^{(i)}, A^{(j)} ).
\label{apoio2}
\end{equation}
Thus, by combining Eqs.~(\ref{apoio1}) and~(\ref{apoio2}) and 
using Eq.~(\ref{Rindler orthonormality}), 
we obtain
\begin{eqnarray}
[\hat{a}_{(\lambda, \varpi, {\bf k}_\bot)},
\hat{a}^{\dagger}_{(\lambda^{\prime}, 
\varpi^{\prime}, {\bf k}_\bot^{\prime})}] 
=
\delta_{\lambda \lambda'} 
\delta(\varpi-\varpi^{\prime})\delta^2
({\bf k}_\bot-{\bf k}_\bot^{\prime}).
\label{comutador de a e a dag}
\end{eqnarray}
The Rindler vacuum 
$\left|0\rangle_{R}\right.$ 
is defined by the condition
\begin{equation}
\hat{a}_{(\lambda,\varpi,{\bf k}_\bot)}\left|0\rangle_{R}\right. = 0.
\nonumber 
\end{equation}

%%%%%%%%%%%%%%%%%%%%%%%%%%%%%%%%%%%%%%%%%%%%%%%%%%%%%%%%%%%%%%%%%%%%%%
\section{The source}
\label{source}
%%%%%%%%%%%%%%%%%%%%%%%%%%%%%%%%%%%%%%%%%%%%%%%%%%%%%%%%%%%%%%%%%%%%%%  
%
Now, let us discuss the source $j^\mu (x)$ to which we will couple the Proca
field $\hat{A}_{\mu} (x)$ as ruled by the interaction Lagrangian density
\begin{equation}
{\cal L}_{\mathrm{int}}= \sqrt{-g}j^{\mu}(x) \hat{A}_{\mu}(x).
	\label{interaction Lagrangian}
\end{equation}
A uniformly accelerated source in Minkowski spacetime with constant 
proper acceleration $a$  corresponds to a static source in Rindler 
coordinates described by the current
\begin{eqnarray}
j^\tau =  q\delta(\xi)\delta(x)\delta(y)\,, \,\,\,\,
j^{\xi}=j^{x}=j^{y}=0\,,
	\label{current}
\end{eqnarray}
where $q$ will play the role of a coupling constant. 
According to the Unruh effect, the inertial vacuum 
corresponds to a thermal state as seen by uniformly accelerated 
observers confined to the Rindler wedge~\cite{CHM-RMP}. Now, 
because the source is static in Rindler coordinates, it can only 
interact with zero-energy Rindler particles and, thus, the 
corresponding spontaneous emission rate must vanish. However, no 
obvious  conclusion can be reached concerning the absorption 
and induced emission rates because of the divergent number of 
zero-energy Rindler particles present in the thermal bath. 
In order to resolve it, we must allow our source to absorb and emit 
(``respond to", for short) nonzero-energy 
particles. This can be achieved by introducing a new parameter, 
$E={\rm const}$, which drives the source to oscillate in time 
and functions as a regulator. Then, in order to keep charge conservation, 
we introduce an extra source at $\xi=L$ to form a dipole with our original
one. The corresponding conserved current associated with the oscillating 
dipole, which will temporarily replace our static source~(\ref{current}), 
can be cast in the form
\begin{eqnarray}
\!\!\!\!\!\! && j^{\tau} 
= \sqrt{2}q\cos(E\tau)[\delta(\xi) -e^{-2aL}\delta(\xi-L)]\delta(x)\delta(y),
\label{regularized conserved current 1}
\\ 
\!\!\!\!\!\! && j^{\xi} = \sqrt{2}q E \sin(E\tau)e^{-2a\xi}
\theta(\xi)\theta(L-\xi)\delta(x)\delta(y), 
\label{regularized conserved current 2}
\\
\!\!\!\!\!\! &&  j^{x} = j^{y}=0. 
\label{regularized conserved current 3}
\end{eqnarray}
Eventually, we take the limits $L\rightarrow+\infty$ and 
$E\rightarrow 0$, and neither the second source at 
$\xi=L$ nor the charge oscillation will contribute to the 
final results. We address to Ref.~\cite{HMS-campo de Maxwell} 
for a more comprehensive discussion on this subject. 

%%%%%%%%%%%%%%%%%%%%%%%%%%%%%%%%%%%%%%%%%%%%%%%%%%%%%%%%%%%%%%%%%%%%%%
\section{Emission and absorption rates in the uniformly accelerated frame}
\label{RF}
%%%%%%%%%%%%%%%%%%%%%%%%%%%%%%%%%%%%%%%%%%%%%%%%%%%%%%%%%%%%%%%%%%%%%%
%  
In this section, we analyze the total response of our charge with respect 
to Rindler observers. Using 
modes~(\ref{complete set of orthonormal solutions 1})-(\ref{complete set of orthonormal solutions 3})
and current~(\ref{regularized conserved current 1})-(\ref{regularized conserved current 3}) 
it is obvious that only modes with $\lambda ={\rm II}$ and~${\rm III}$  can couple to the charge.
We focus first on the emission rate of Rindler particles with $\lambda = {\rm II}$.

At the tree level, the emission amplitude of Rindler particles with 
quantum numbers $({\rm II},\varpi,{\bf k}_\bot)$ out to the
Rindler vacuum is
\begin{equation}
{\cal A}_{({\rm II}, \varpi,{\bf k}_\bot)}^{\mathrm{e}\mathrm{m}}=_
{R}\langle {\rm II},\varpi,{\bf k}_\bot| 
i \int d^{4}x {\cal L}_{\mathrm{int}}|0\rangle_{R}.
\nonumber
\end{equation}
The corresponding differential emission rate  
is given by
\begin{equation}
dW_{0}^{\rm em}({\rm II},\varpi,{\bf k}_\bot)=
|{\cal A}_{({\rm II},\varpi,{\bf k}_\bot)}^{\rm em}|^{2}
 d\varpi \, d^2 {\bf k}_\bot/T,  
\nonumber
\end{equation}
where $T$ is the (arbitrarily large) proper time interval during 
which the interaction remains turned on.
Taking, at this point, the limit $ L\to+\infty$  
to eliminate the influence of the extra source on 
$dW_{0}^{\rm em}({\rm II},\varpi,{\bf k}_\bot)$, we obtain         	
\begin{eqnarray}
&& dW_{0}^{\rm em}(
{\rm II},\varpi,{\bf k}_\bot)= \frac{q^{2}}{4\pi^{3}a}\sinh(\pi E/a) \delta(\varpi-E)
 \nonumber \\
&& \times\, \left| K_{iE/a}^{\prime} \left( \frac{\rho}{a} \right) 
 + {\cal O} (E) \right|^2  
\,  d\varpi \, d^2  {\bf k}_\bot, 
\label{dW0em}
\end{eqnarray}
as $E \to 0 $.
(Note that charges at $\xi = L \to +\infty$ are inertial and, thus, 
become ``inert" for our present purposes.)
Now, we should remember that according to the Unruh effect 
the Minkowski vacuum  corresponds to a thermal bath of Rindler 
particles at a temperature $\beta^{-1}=a/(2 \pi)$. 
Then, the differential emission rate of massive vector Rindler 
particles with $\lambda={\rm II}$ into the Unruh thermal bath for 
fixed ${\bf k}_\bot$ according to uniformly accelerated observers is
\begin{equation}
P_{{\rm II},{\bf k}_\bot}^{\rm em} d^2 {\bf k}_\bot
= 
\int dW_{0}^{\rm em}
({\rm II},\varpi,{\bf k}_\bot) \left(1 + \frac{1}{e^{2\pi\varpi/a}-1} \right),
\label{PemIIKbot}
\end{equation}
where the first and second terms inside the parenthesis are associated 
with spontaneous and induced emission, respectively, and the integration is
performed only in the $\varpi$ variable.
By evaluating the integral above and taking the limit $E\rightarrow 0$ 
at the end, we obtain
\begin{equation}
P_{{\rm II},{\bf k}_\bot}^{\rm em}d^2 {\bf k}_\bot =
\frac{q^{2}}{8\pi^{3}a}|K_{1}(\rho/a)|^{2} d^2 {\bf k}_\bot.
\label{PIIem}
\end{equation}
The necessity for the introduction of the regularization
parameter $E$ can be appreciated from Eq.~(\ref{PemIIKbot})
by noticing that the term that multiplies $\delta (\varpi -E)$
in the right hand side of Eq.~(\ref{dW0em}) vanishes as 
$\varpi \sim E\to 0$ while the Planckian (induced emission)  
term in Eq.~(\ref{PemIIKbot}) diverges as $\varpi \to 0$.
Analogously, we can obtain the differential absorption 
rate:
\begin{equation}
P_{{\rm II},{\bf k}_\bot}^{\rm abs} d^2 {\bf k}_\bot
= 
\int dW_{0}^{\rm abs}
({\rm II},\varpi,{\bf k}_\bot) \frac{1}{e^{2\pi\varpi/a}-1}.
\label{PabsIIKbot}
\end{equation}
 We see that
\begin{equation}
P_{{\rm II},{\bf k}_\bot}^{\rm abs} d^2 {\bf k}_\bot=
P_{{\rm II},{\bf k}_\bot}^{\rm em} d^2 {\bf k}_\bot 
\label{PIIabs}
\end{equation}
because the spontaneous emission becomes 
negligible in comparison with the induced emission as 
$E$ vanishes.

Next, we must repeat the same calculations for 
$\lambda = {\rm III}$. It turns out that 
$
P_{{\rm III},{\bf k}_\bot}^{\rm em} d^2 {\bf k}_\bot= 
P_{{\rm III},{\bf k}_\bot}^{\rm abs} d^2 {\bf k}_\bot = 0.
$
This is so because 
$$
dW_0^{\rm em} ({\rm III}, \varpi, {\bf k}_\bot) 
\sim  
{\cal O} (\varpi^2) \delta (\varpi -E) d\varpi d^2 {\bf k}_\bot
$$ 
as 
$\varpi \sim E \to 0$, while $1/(e^{2\pi \varpi/a}-1) \sim {\cal O} (\varpi^{-1})$
as $\varpi \to 0$ 
[and analogously for $dW_0^{\rm abs} ({\rm III}, \varpi, {\bf k}_\bot)$].
As a result, only mode II contributes to the total differential response rate: 
\begin{eqnarray}
P_{{\bf k}_\bot}^{\mathrm{tot}} d^2 {\bf k}_\bot 
&& 
=
(
P^{\rm em}_{{\rm II},{\bf k}_\bot}
+
P^{\rm abs}_{{\rm II},{\bf k}_\bot}
) 
d^2 {\bf k}_\bot
\nonumber \\
&&
=
\frac{q^{2}}{4\pi^{3}a}|K_{1}(\rho /a)|^{2} d^2 {\bf k}_\bot.
\label{resposta em Rindler}
\end{eqnarray}

%%%%%%%%%%%%%%%%%%%%%%%%%%%%%%%%%%%%%%%%%%%%%%%%%%%%%%%%%%%%%%%%%%%%%%%%%%%%%%%%%
\section{Emission rate in the inertial frame}
\label{IF}
%%%%%%%%%%%%%%%%%%%%%%%%%%%%%%%%%%%%%%%%%%%%%%%%%%%%%%%%%%%%%%%%%%%%%%%%%%%%%%%%%

Now, let us compute the response rate for  a uniformly accelerated charge as 
given by Eq.~(\ref{current}) coupled to the Proca field through 
Eq.~(\ref{interaction Lagrangian}) with respect to inertial observers.
These observers cover the Minkowski spacetime with Cartesian coordinates, 
in which case the metric components assume the usual form
$\eta_{\mu \nu}={\rm diag} \, (1,-1,-1,-1)$.
Hence, it is convenient to write the components of 
current~(\ref{current}) in the Cartesian basis as 
\begin{eqnarray}
j^{t} &=& q a z\delta(x)\delta(y)\delta(\xi),
\label{j no ref iner 1} 
\\ 
j^{z} &=& q a t\delta(x)\delta(y)\delta(\xi),
\label{j no ref iner 2} 
\\
j^{x}&=&j^{y}=0, 
\label{j no ref iner 3}
\end{eqnarray} 
where $\delta (\xi) = \delta(z-\sqrt{t^2+ a^{-2}})/(a \sqrt{t^2 + a^{-2}})$.

An orthonormal set of solutions for Eq.~(\ref{Lagrange equations})
proportional to plane waves and satisfying Eq.~(\ref{Lorenz constraint})
can be cast in the form
$$
A_{\mu} ({\bf k}, \lambda) = 
(16 \pi^3 \omega)^{-1/2} \epsilon_{\mu}({\bf k},\lambda) e^{i(- \omega t + {\bf k}\cdot {\bf x} )}
$$
with polarization vectors 
\begin{eqnarray}
\label{vetor 1}
&&\epsilon^{\mu}({\bf k},1) = (0,{\hat \epsilon}({\bf k},1)), \\
&&\epsilon^{\mu}({\bf k},2) = (0,{\hat \epsilon}({\bf k},2)), \\
&&\epsilon^{\mu}({\bf k},3) =  ({|{\bf k}|}/{m},{\omega}\hat{k}/{m})
\label{vetor 4}
\end{eqnarray}
where ${\hat \epsilon}({\bf k},1)$ and ${\hat \epsilon}({\bf k},2)$ 
are unit three-vectors orthogonal to each other and to   
${\bf k} \equiv (k_x,k_y,k_z)$, 
$\hat k \equiv {\bf k}/|{\bf k}|$, 
and 
$\omega \equiv \sqrt{m^2 + |{\bf k}|^2}$. 
(In this case, the nonphysical mode orthogonal to the physical 
modes above, satisfying Eq.~(\ref{Lagrange equations}) but 
not Eq.~(\ref{Lorenz constraint}) has polarization 
$\epsilon^{\mu}({\bf k},4) = k^\mu/m$.)
Next, by using the Klein-Gordon inner product~(\ref{KG}) 
(where in this case the Cauchy surface $\Sigma$ is given 
by any $t= {\rm const}$ hypersurface), we verify that the
physical modes satisfy 
\begin{eqnarray} 
(A^{(i)},A^{(j)})= \delta_{\lambda \lambda'} \delta^3 ({\bf k}-{\bf k}^{\prime}),
\label{matriz no ref inercial}
\end{eqnarray}
where $(i),(j)$ represent the quantum numbers $({\bf k}, \lambda)$ 
($\lambda= 1,2,3$).

The Proca field in the inertial frame is expanded, thus, as 
\begin{eqnarray}
\hat{A}_{\mu} (x^\nu) = \int d^3 {\bf k}\sum_{\lambda = 1}^{3}
[\hat{a}_{({{\bf k},\lambda})} A_{\mu}({\bf k},\lambda) + {\rm H.c.} ].
\label{exp no ref inercial}
\end{eqnarray}
It is straightforward to verify that $\hat{A}_{\mu}$ and the canonically 
conjugate momentum  $\hat{\Pi}^{\mu}$ satisfy canonical commutation
relations provided that (see, e.g., Ref.~\cite{GreinerR})
\begin{equation}
[\hat{a}_{({\bf k},\lambda) },
\hat{a}^{\dagger}_{({\bf k}',\lambda')}] 
=  
\delta_{\lambda \lambda'}
\delta^3 ({\bf k} - {\bf k}^{\prime}).
\label{comutator between a and a dag}
\end{equation}

Now, because inertial observers experience no particles in the 
Minkowski vacuum $|0\rangle_{M}$ [defined by 
$\hat{a}_{({\bf k},\lambda) } |0\rangle_{M} \equiv 0$ for all $({\bf k},\lambda)$], 
no absorption processes are allowed according
to them. As a result, only the emission rate will contribute to the 
total response. At the tree level, the emission amplitude of a Proca 
particle with three-momentum ${\bf k}$ and 
polarization $\lambda$ as defined by inertial observers 
is
\begin{eqnarray*}
^{\rm M}{\cal A}^{\rm em}_{({\bf k},\lambda)}
 =  _{M}\langle {\bf k},\lambda|
i\int d^4x{\cal L}_{\mathrm{int}}|0\rangle_{M}.
\end{eqnarray*}
The differential emission rate of Minkowski Proca particles with fixed transverse 
momentum  ${\bf k}_\bot$ is given by
\begin{eqnarray}
^{\rm M}P^{\rm tot}_{{\bf k}_\bot} d^2 {\bf k}_\bot = d^2 {\bf k}_\bot
\sum_{\lambda = 1}^{3}
\int_{-\infty}^{+\infty}dk_{z} {|^{\rm M} {\cal A}^{\rm em}_{({\bf k},\lambda)}|^{2}}/{T}.
\label{PM1}
\end{eqnarray}
Next, we use the identity 
$$
\sum_{\lambda = 1}^{3}\epsilon_{\mu}({\bf k},
\lambda)\epsilon_{\nu}({\bf k},\lambda) = k_\mu k_\nu/m^2 - \eta_{\mu\nu}
$$ 
satisfied by the  polarization vectors to cast Eq.~(\ref{PM1}) in the form
\begin{eqnarray*}
&& ^{\rm M}P^{\rm tot}_{{\bf k}_\bot}d^2 {\bf k}_\bot 
=
d^2 {\bf k}_\bot \int_{-\infty}^{+\infty}dk_{z}
\int \frac{d^4xd^4x^{\prime}}{16 \pi^3\omega} 
\nonumber \\
&& \times  
j^{\mu}(x) j^{\nu}(x^{\prime})
\left( \frac{k_{\mu}k_{\nu}}{m^2} -\eta_{\mu\nu}\right)
e^{i\omega(t - t^{\prime})}e^{-i{\bf k}\cdot ({\bf x}-{\bf x}^{\prime})}.
\nonumber
\end{eqnarray*}
Now, by using current conservation $\partial_\mu j^\mu=0$, we can show that 
terms in the integral proportional to $j^{\mu} k_{\mu}$ do not contribute. 
Finally, by using Eqs.~(\ref{j no ref iner 1})-(\ref{j no ref iner 3}) we find 
\begin{eqnarray}
^{\rm M}P^{\rm tot}_{{\bf k}_\bot} d^2 {\bf k}_\bot 
= 
\frac{q^{2}}{4\pi^{3}a}
\left|K_{1}\left(\rho/a\right)\right|^{2} 
d^2 {\bf k}_\bot.
\label{resposta em Minkowski}
\end{eqnarray} 
The equality between Eqs.~(\ref{resposta em Minkowski}) 
and~(\ref{resposta em Rindler}) can be interpreted in 
terms of known elementary particles as follows. Firstly, 
let us note that the Lagrangian density~(\ref{interaction Lagrangian})
can be used to describe the interaction of noninertial 
electrons emitting $Z^0$ bosons. Then, 
the equality between Eqs.~(\ref{resposta em Rindler}) 
and~(\ref{resposta em Minkowski}) reflects, in this
case, the fact that each (nonzero energy) $Z^0$ emitted 
from a uniformly accelerated $e^-$  in the Minkowski vacuum 
as described by inertial observers corresponds to either 
the emission to or absorption from the Unruh thermal 
bath of a zero-energy $Z^0$ Rindler boson as described 
by coaccelerated observers.

%%%%%%%%%%%%%%%%%%%%%%%%%%%%%%%%%%%%%%%%%%%%%%%%%%%%%%%%%%%%%%%%%%%%%%%%
\section{Concluding remarks}
\label{Conclusions}
%%%%%%%%%%%%%%%%%%%%%%%%%%%%%%%%%%%%%%%%%%%%%%%%%%%%%%%%%%%%%%%%%%%%%%%%

We have canonically quantized the Proca field in the Rindler wedge.
The results obtained were applied to compute the total response rate of a 
source with constant proper acceleration interacting with the Proca field
from the point of view of observers coaccelerated with the source. 
We have shown that this corresponds to the combined emission 
and absorption rates of zero-energy Proca particles to and from
the Unruh thermal bath, respectively, and that only Rindler  particles 
with $\lambda = {\rm II}$ contribute. Then, we have explicitly verified 
that the total response rate obtained in the Rindler wedge is exactly the 
same as the total emission rate of Minkowski particles as described
by inertial observers. This result is consistent with the fact that 
each particle emitted in the inertial frame must correspond to {\it either} 
the emission {\it or} absorption of a Rindler particle in the accelerated 
frame, since both observers must agree concerning 
changes in the state of the quantum field. Finally, because of the close 
conceptual relationship between the Unruh thermal bath and Hawking radiation, 
we can forecast that the total response of a static source in the vicinity 
of a black hole coupled to the Proca field will be well approximated by 
$
\int P_{{\bf k}_\bot}^{\mathrm{tot}} d^2 {\bf k}_\bot 
$,
where $P_{{\bf k}_\bot}^{\mathrm{tot}} d^2 {\bf k}_\bot$ is given in 
Eq.~(\ref{resposta em Rindler}) with $a$ being identified with the 
proper acceleration of the static source outside the black hole. 

%%%%%%%%%%%%%%%%%%%%%%%%%%%%%%%%%%%%%%%%%%%%%%%%%%%%%%%%%%%%%%%%%%
\acknowledgments
%%%%%%%%%%%%%%%%%%%%%%%%%%%%%%%%%%%%%%%%%%%%%%%%%%%%%%%%%%%%%%%%%%
%
The authors are grateful to Conselho Nacional de Desenvolvimento 
Cient\'\i fico e Tecnol\'ogico  and to Coordena\c{c}\~ao de 
Aperfei\c{c}oamento de Pessoal de N\'\i vel Superior for partial 
financial support. J.C., E.C., and L.C also acknowledge
Funda\c{c}\~ao de Amparo \`a Pesquisa do Estado do Par\'a 
while G.M. is grateful to Funda\c c\~ao de Amparo \`a Pesquisa 
do Estado de S\~ao Paulo for partial support. We are indebted to 
Atsushi Higuchi for various discussions.

\end{document}